\documentclass[twocolumn,showpacs,preprintnumbers,amsmath,amssymb]{revtex4}

\usepackage{graphicx}
\usepackage{dcolumn}
\usepackage{bm}

\begin{document}


\title{Electronic spin drift in graphene field effect transistors}

\author{C. J\'{o}zsa$^1$}
 \email{C.Jozsa@rug.nl}
 \author{M. Popinciuc$^2$, N. Tombros$^1$, H.T. Jonkman$^2$}
\author{B.J. van Wees$^1$}%

\affiliation{$^1$Physics of Nanodevices, $^2$Molecular
Electronics, Zernike Institute for Advanced Materials, University
of Groningen, The Netherlands
}%

\date{\today}

\begin{abstract}

We studied the drift of electron spins under an applied DC
electric field in single layer graphene spin valves in a field
effect transport geometry at room temperature. In the metallic
conduction regime ($n \simeq 3.5 \times 10^{16}$~m$^{-2}$), for DC
fields of about $\pm$70~kV/m applied between the spin injector and
spin detector, the spin valve signals are increased/decreased,
depending on the direction of the DC field and the carrier type,
by as much as $\pm$50\%. Sign reversal of the drift effect is
observed when switching from hole to electron conduction. In the
vicinity of the Dirac neutrality point the drift effect is
strongly suppressed. The experiments are in quantitative agreement
with a drift-diffusion model of spin transport.
\end{abstract}

\maketitle

Graphene, a two-dimensional hexagonal carbon lattice, was isolated
from graphite and became available for electric transport type
measurements in 2004 \cite{Nov2004,RiseofG}. Graphene is unique in
its nature by allowing the experimentalist to shift the Fermi
level trough the zero-gap energy bands and continuously tune the
carrier density from holes to electrons via the Dirac neutrality
point. Recent studies also addressed the spin degree of freedom,
resulting in successful injection, transport and detection of
spins in single and multiple layers of graphene
\cite{Tombros2007,Hill2006,Cho2007,Nishioka2007,Ohishi2007,Kawakami2008}. The
spin relaxation time $\tau_{sf}$ measured about 150 ps (as
extracted from Hanle type spin precession measurements), which
combined with a diffusion constant of $D\simeq0.02~$m$^2$/s
yielded a considerable room temperature spin diffusion length
$\lambda_{sf}=\sqrt{D\tau_{sf}}$ of 1.5 to 2~$\mu$m.

\begin{figure}[b!]
\includegraphics[width=8.5cm]{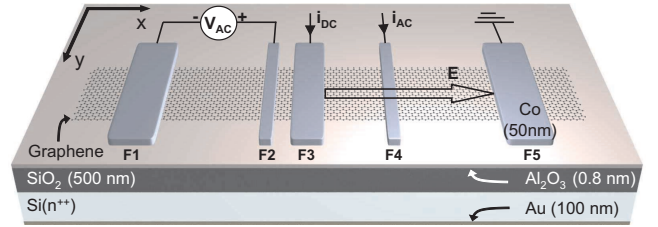}
\caption{\label{fig:device} Schematic drawing of the spin drift
experiment. The Co contact widths are 350~nm (for F1), 90~nm (F2),
250~nm (F3), 90~nm (F4), 350~nm (F5) and the distances between them
are (from left to right) 2~$\mu$m, 0.6~$\mu$m, 1.3~$\mu$m and
2~$\mu$m. The $0.9 \times 7$~$\mu$m$^2$ sized graphene flake is
covered by an aluminum oxide layer. A (gate) voltage, applied
between the Au contacted Si(n$^{++}$) substrate and the grounded
electrode F5, allows for the control of the charge density in
graphene. The layer thicknesses are given in the brackets.}
\end{figure}

Until now, spin transport in graphene was studied in the low bias
regime where the transport is described by diffusion only. In this
paper we study the spin transport in the presence of DC electric
fields. Due to the large carrier mobilities in graphene 
\cite{KimUltrahighMobility2008}, the DC fields give rise to spin 
drift effects which are comparable in
magnitude to the spin diffusion in graphene and, therefore, become
accessible experimentally. In order to study the spin drift effect,
we have built lateral spin valve devices as illustrated in
Fig.~\ref{fig:device}. The graphene layer is contacted by a series
of Co electrodes (F1-F5) used for the injection/detection of spins
and the application of the DC electric fields. The dimensions of the
contacts and their spacings given in the figure caption are the ones
of the sample discussed in this paper and are characteristic for the
spin valves we have fabricated. The graphene flakes we use are
typically between 0.2 and 4~$\mu$m wide and up to 50~$\mu$m long.
Besides the AFM thickness characterization of every flake, on
several samples we have conducted Raman spectroscopy measurements as
well \cite{Fer2006Raman}, in order to select the single layer
flakes. For a detailed description of the device fabrication see
Ref.~\onlinecite{Tombros2007}. The contact resistances are between
2.5~k$\Omega$ and 12~k$\Omega$. A standard low-frequency AC lock-in
technique with a current of 4~$\mu$A RMS was used for measuring the
spin signals. The measurements were performed at room temperature
and in vacuum in order to minimize the shifts of the Dirac point due
to charging effects of adsorbates on the graphene layer.

\begin{figure}[t!]
\includegraphics[width=8.5cm]{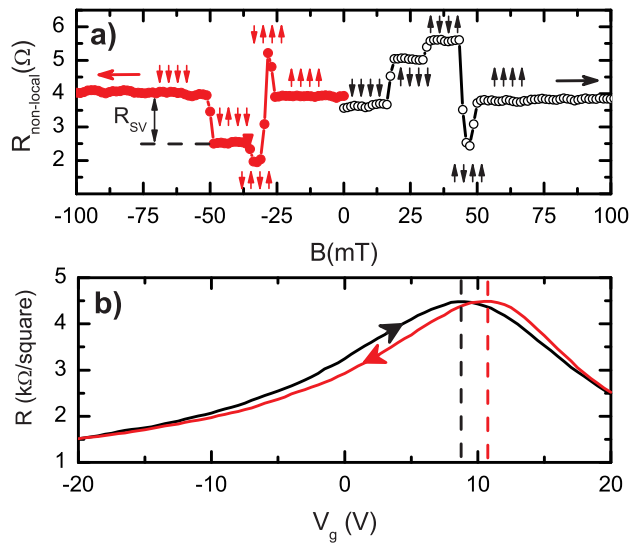}
\caption{\label{fig:SV} a) Non-local spin valve measurement of the
device presented in Fig.~\ref{fig:device}. The small arrows
represent the magnetic orientation of electrodes F1, F2, F4 and
F5. The asymmetric spin valve behavior is due to the switching
order of F4 and F5 which depends on external field sweep direction
(indicated by the large horizontal arrows). $R_{SV}$ represents
the spin signal; see text for details. b) 4-point measurement of
the graphene resistivity between contacts F3 and F4 versus the
gate voltage.}
\end{figure}

An AC spin imbalance is created in graphene by
injection/extraction of spin currents by the ferromagnetic
electrodes F4 and F5 through an aluminium-oxide tunnel barrier.
Due to the opaque Co/Al$_2$O$_3$/graphene interface, the spin
transport is not disturbed by the central contact F3. The spin
imbalance diffuses symmetrically, with respect to the injection
points F4 and F5, through the device giving rise to an AC voltage
drop between the ferromagnetic detector electrodes F2 and F1. In
this non-local measurement geometry, the spin transport is thus
completely separated from the charge transport and is solely
described by the diffusion constant $D$ and the spin relaxation
time $\tau_{sf}$. The non-local voltage drop between F2 and F1 is
measured as a function of a magnetic field applied along the easy
axis of the Co electrodes (y direction). This results in typical
spin valve signals as shown in Fig.~\ref{fig:SV}~a). The variable
width of the Co electrodes yields a difference in their coercive
fields (20 to 50 mT in the present case), allowing for separate
switching in the external magnetic field. Contacts with similar
widths, however, can sometimes interchange in their switching
order depending on, e.g., the domain wall nucleation, on the
sweeping direction of the field (positive to negative values and
vice versa). Such an effect is observed in the measurement shown
in Fig.~\ref{fig:SV}~a) where the switching order of F4 and F5 is
interchanged on the negative versus positive field's side. The
identification of the switches could nevertheless be done by
examining the spin valve signals obtained in different regions of
the same sample by choosing different pairs of injector/detector
electrodes.

Each discrete resistance level seen in Fig.~\ref{fig:SV}~a)
corresponds to a combination of magnetically parallel/antiparallel
electrodes. The step heights carry information about the importance
of that specific electrode's contribution to the spin
injection/detection process. The highest change in signal is
observed when the inner electrodes (F2,F4) change their relative
parallel/antiparallel magnetization state. The further away an
electrode is, the smaller its corresponding resistance step is. The
spins injected at F5 must diffuse longer (and have thus higher
chance to relax) before they are detected by the detectors (F2,F1)
than the spins injected at F4. The presence of four switches in our
measurements proves the spin transport through the graphene channel
under all magnetic electrodes, over the full $\simeq 6~\mu$m length
of the device.

Spin drift effects in all-electrical n-GaAs based devices were measured by
Lou et al. \cite{LouGaAs2006}, where the conduction was limited to
electrons. Unlike in semiconductors, in graphene, due to its band structure,
electrostatic gating allows for switching from hole to electron
conduction while keeping the carrier mobility, diffusion constant,
Fermi velocity, electric conductivity and other parameters
approximately unchanged. The gating effect is reflected in the
graphene resistivity as plotted in Fig.~\ref{fig:SV}b). The
position of the Dirac neutrality point separating hole and
electron conduction regimes, as identified by the minimum in
conductance \cite{Nov2005Dirac,Tan2007Dirac}, shows a small
hysteresis (+~9~V to +~11~V) for the two sweep directions. The
measurement also reveals a carrier mobility of
$\mu=$0.25~m$^2$/Vs.

Applying a DC electric field to generate carrier and spin drift in
addition to the diffusion would in principle be possible by
passing a DC current from F2 to F4. However, a strong DC bias can
dramatically influence the spin injection efficiency of the
electrodes \cite{YuFlatte2,NatPhys} making it impossible to
separate the drift effect contribution. We have conducted
experiments on graphene spin valve devices where both AC spin
injector electrodes are DC biased and we measured changes of the
spin valve effect up to an order of magnitude compared to the
non-biased measurements \cite{JozsaEfficiency}. In order to avoid
the manipulation of the spin injection efficiency, we have
introduced the electrode F3 (see Fig.\ref{fig:device}). The DC
field is applied between the electrodes F3 and F5 by sending a
constant DC current between them, with F5 being grounded. We could
not avoid sharing the drain of the AC and DC circuits, the
injector F5 on the far right of the device is effectively DC
biased. However, the effect of changing its injection efficiency
with the DC current bias is minimized as discussed later.

Experimentally, the drift will manifest itself as an asymmetric
distribution of the spin imbalance on the two sides of the spin
injection point F4 (and F5), in addition to the symmetric diffusion
process. Between F3 and F5 thus, where the field E is active, a
favored direction of the spin transport will be defined by the
electric field's polarity and the carrier type (electrons or holes).
Such a device with fixed-distance injector/detector electrodes will
therefore show a variable spin signal dependent on the DC electric
field.

\begin{figure}[b!]
\includegraphics[width=8.5cm]{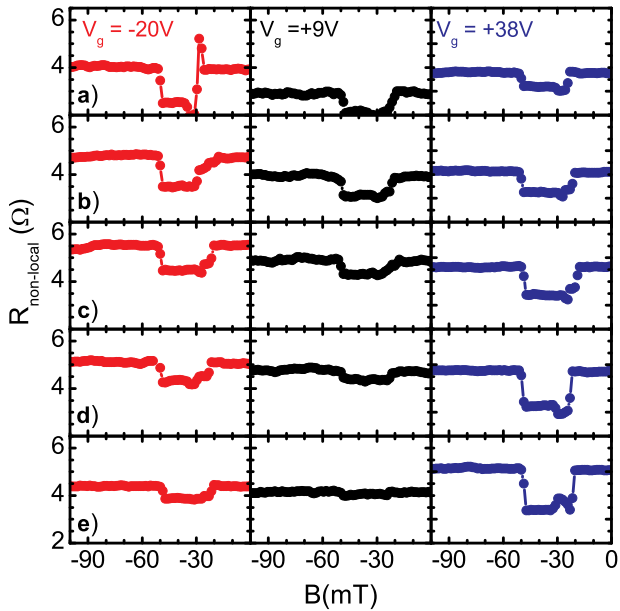}
\caption{\label{fig:meas} The drift effect in a graphene spin
valve. The gate voltage V$_g$ varies from left to right and it is
indicated for each column. The DC electric current varies from top
to down as: (a) I$_{DC}$ = -40 $\mu$A, (b) -20 $\mu$A, (c) 0
$\mu$A, (d)+20 $\mu$A, (e) +40 $\mu$A. For clarity only the
negative side of the spin valve measurements are plotted. }
\end{figure}

In Fig.~\ref{fig:meas} we present room temperature spin valve
measurements for several DC currents (applied between F3 and F5)
at three gate voltages: -20~V (hole conduction), +9~V (the Dirac
neutrality point) and +38~V (electron conduction). The upper limit
of the electric fields was set by the current densities allowed to
pass through the tunnel barriers without damaging them\footnote{We
have sent DC currents as high as 80 $\mu$A through a similar
device, at which point the Co/Al$_2$O$_3$ electrodes were
damaged.}. Before each measurement, the magnetic field is ramped
up to 300~mT to saturate the magnetization of the contact
electrodes. The DC currents used were $\pm$40, $\pm$20 and 0
$\mu$A and their corresponding electric fields were $\pm$68~kV/m,
$\pm$34~kV/m for V$_g$=-20 and +38~V and $\pm 206$~kV/m, $\pm
103$~kV/m for V$_g$=+9~V. Similar measurements, done on a set of
samples with different geometry and/or contact resistances,
support the data we present and discuss here.

The change in the injection efficiency of the electrode F5 (common
ground for the DC and AC currents) for different DC fields appears
in the spin valve measurements as an increased/decreased
contribution of this electrode to the device non-local resistance.
This can be seen in Fig.~\ref{fig:meas} where, for the
measurements performed at +38~V gate voltage, the resistance step
at approximately -30~mT (corresponding to the magnetic switch of
electrode F5) does show a dependence on the DC electric field.
However, this electrode is positioned far from the AC detection
circuit, and we can minimize its effect by defining the "spin
signal" $R_{SV}$ as the resistance step between the parallel and
antiparallel magnetic state of the two \emph{central} electrodes
F2 and F4 (see Fig.~\ref{fig:SV}). Therefore, the DC field
dependence of $R_{SV}$ indicates the presence of carrier drift as
explained below.

At -20~V gate voltage (left column of graphs in
Fig.~\ref{fig:meas}), we are in the high density hole conduction
regime with a carrier concentration of $n_h \simeq 3.5 \times
10^{16}$~m$^{-2}$ yielding a graphene resistivity $R_{square}
\simeq1.5$ k$\Omega$. The non-local spin signal $R_{SV}$ reads
1~$\Omega$ when no DC field is present (graph c). Applying a DC
field of +34~kV/m (graph d) and +68~kV/m (graph e) decreases the
spin valve signal to 0.7~$\Omega$ and 0.45~$\Omega$ respectively,
indicating spin drift in the region F3-F5 that counteracts the
symmetric spin diffusion. This transport is labeled as "upstream"
since, in order to detect a spin valve signal at electrodes F2 and
F1, the spins need to travel against the action of the carrier
drift. Now, if the DC electric fields' polarity is reversed, i.e.
we apply a field of -34~kV/m (graph b) and -68~kV/m (graph a), the
resulting drift facilitates spin transport towards the detectors
("downstream" transport) and the measured spin valve signal
increases to 1.3 and 1.5 $\Omega$ respectively.

In the case of electron conduction, the drift effect reverses,
because the electric field - charge carrier interaction is opposite,
as seen in the set of curves measured at V$_g$=+38~V (right column
in Fig~.\ref{fig:meas}). This voltage was selected to give
approximately the same parameters ($n_e, R_{local}$) on the electron
conduction side of the Dirac neutrality point as for the hole
conduction (V$_g$=-20~V). The spin signal scales with the field
amplitude and direction as expected, reading 0.6, 0.9, 1.2, 1.5 and
1.7~$\Omega$ for -68, -34, 0, +34 and +68~kV/m respectively.

At a gate voltage that locates the Fermi level close to the Dirac
neutrality point ($R_{square}\simeq4.5$~k$\Omega$) the conduction
happens neither via electrons, nor via holes, which means that no
spin drift should take place. Examining the measurements at
V$_g$=+9~V (central column of graphs in Fig.~\ref{fig:meas}), we
observe a slight decreasing tendency of the spin valve signal for
increasingly positive DC fields, similar to the drift in case of
hole conduction. This behavior can be explained by the interplay
of the following two effects.

First, under the high DC current a certain shift of the effective
gate voltage is expected. Considering the 20~k$\Omega$ resistance of
the structure (graphene + contact F5), this shift equals to
$\pm$0.8~V for currents of $\pm$40~$\mu$A. Unlike in the case of
high carrier densities, around the neutrality point such a shift can
critically change the carrier density in graphene and lead to spin
drift. Second, due to a shift of the Dirac point in time under a
constant gate voltage stress, it is difficult to set and keep a gate
voltage just at the right position to assure neutrality. In our
case, it seems that we were slightly in the hole conduction regime
when measuring at V$_g$=+9~V.

\begin{figure}[h!]
\includegraphics[width=8.5cm]{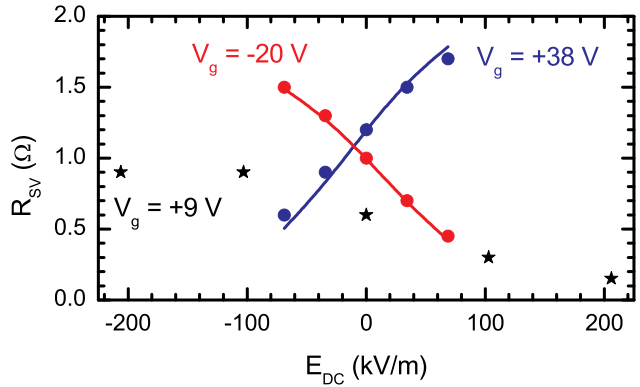}
\caption{\label{fig:simu} Spin valve signal for high carrier
densities (circles), and near the Dirac neutrality point (stars).
The lines represent a calculation (with no free parameters) of the
drift effect based on the drift-diffusion equation.}
\end{figure}

For a quantitative interpretation of the data we have extracted the
spin signal values from all the measurements of Fig.~\ref{fig:meas}
and plotted them in Fig.~\ref{fig:simu} against the DC field for the
three gate voltages. For the theoretical description, we adopt the
drift-diffusion model introduced by Yu and Flatt\'{e}
\cite{YuFlatte1,YuFlatte2}. In steady-state, the drift-diffusion
equation for the spin imbalance $n_s$ reads:
\begin{equation}\label{Eq:DriftDiff}
D \nabla^2 n_s + \mu E \nabla n_s - \frac{n_s}{\tau_{sf}}=0,
\end{equation}
where $D$ is the diffusion constant, $\tau_{sf}$ is the spin
relaxation time, $\mu$ is the carrier mobility and $E$ is the DC
electric field. The term $\mu E$ represents the drift velocity
$v_d$ and its magnitude compared to the Fermi velocity $v_F =
10^6$~m/s is important for the contribution of the spin drift to
the spin transport. Similar to the spin diffusion length
$\lambda_{sf} = \sqrt{D\tau_{sf}}$, we define a spin drift length
$\lambda_d = D/v_d$. The symmetric diffusion and asymmetric drift
effects add up to form a spin transport characterized by a pair of
length scales, named \emph{upstream} and \emph{downstream} lengths
$\lambda_{\pm}$ \cite{YuFlatte1}:
\begin{equation}\label{Eq:UpDownStream}
\frac{1}{\lambda_{\pm}} = \pm \frac{1}{2} \frac{1}{\lambda_d} +
\sqrt{\frac{1}{4} \frac{1}{\lambda_d^2} + \frac{1}{\lambda_{sf}^2}
}.
\end{equation}
The general solution to the spin imbalance equation
\ref{Eq:DriftDiff} in the direction $x$ parallel to the spin
transport is
\begin{equation}\label{Eq:Sol}
n_s(x) = A \exp{(+\frac{x}{\lambda_+})} +
B\exp{(-\frac{x}{\lambda_-})},
\end{equation}
where A and B are determined by the boundary conditions. Since we
take the electrode F4 as the spin source, only the decaying
solution with respect to the injection point need to be
considered. Therefore, we can write the spin signal ($R_{SV}$)
dependence versus the applied DC field in the form:
\begin{equation}\label{Eq:Rdependence}
R_{SV} = R_0 \exp{(-\frac{L}{\lambda_{\pm}})},
\end{equation}
where $R_0$ is the spin signal that would be measured at the
injection point, $\lambda_{\pm}$ carries the electric field
dependence (Eq.~\ref{Eq:UpDownStream}) and $L \simeq 1.5\mu$m is
the distance between F3 and F4.

In the following we compare the drift-diffusion equation for the
spin signal (Eq.~\ref{Eq:Rdependence}) with the spin drift
measurements in the high carrier density regime. From the
conductivity measurements presented in Fig.~2b we extract $\mu
\simeq 0.25 $~m$^{2}$/Vs using $\sigma=ne\mu$, where the carrier 
concentration $n$ is calculated from the sample's response to gate 
voltage (fig. 2b), using the method described in 
Refs.~\onlinecite{RiseofG},\onlinecite{Tan2007Dirac}. This gives a 
drift velocity of $v_d \simeq \pm 1.7 \cdot 10^4$ m/s
for $E \simeq \pm 68$~kV/m. The diffusion constant of $D \simeq
0.02$~m$^2$/s is obtained using the Einstein relation
$\sigma=Ne^2D$, where the density of states $N$ is determined
according to Ref.~\onlinecite{Tombros2007}. Using
Eq.~\ref{Eq:UpDownStream} and a spin relaxation length
$\lambda_{sf} \simeq 2 \mu$m \cite{Tombros2007}, we calculate the
magnitude of the drift effect manifesting in the up/downstream
lengths as $\lambda_+ = 1.3 \mu$m, $\lambda_- = 3.0 \mu$m for E
=$\pm34$~kV/m and $\lambda_+ = 0.9 \mu$m, $\lambda_- = 4.4 \mu$m
for E =$\pm68$~kV/m. Note, that at E =$\pm68$~kV/m the
characteristic length scale of the drift-diffusion spin transport
is about two times increased/decreased compared to the diffusion
process. The value of $R_0$ we estimate from the spin valve
measurements in zero electric field when
$\lambda_{\pm}=\lambda_{sf}$ and $R_0$ is given by
$R_0=R_{SV}\exp{(L/\lambda_{sf})}$. By taking the length of the
drift region $L=1.5\mu m$, $\lambda_{sf}=2\mu m$ and the measured
$R_{SV}$ at E =0 we can determine $R_0$. Now, we are able to
calculate the drift effect according to Eq.~4, all parameters
being known. The theoretical curves (solid lines in
Fig.~\ref{fig:simu}) are in excellent agreement with the
experimental data. There are no free parameters in the calculation
and, in fact, these measurements could be used to extract
$\lambda_{sf}$, if it would be unknown.

At the Dirac point, due to the symmetry, no drift effect is
expected. However, in our measurements we are probably slightly
away from the Dirac point. Here we cannot apply our above
analysis. It remains an open theoretical question to what the
drift velocity should be at the charge neutrality point.

In conclusion, we studied spin transport manipulation in graphene,
by carrier drift under the action of a DC electric field. For high
charge carrier densities ($n \simeq 3.5 \times 10^{16}$~m$^{-2}$)
depending on the direction of the applied DC field and the nature
of the carriers we were able to modify the effective spin
relaxation length by factor of 4.8 which resulted in a modulation
of the spin valve signal of about 300\%. The spin-drift
measurements are described well by a drift-diffusion model. The
control over the drift velocity we demonstrated opens new
possibilities for exploring spin related phenomena in other types
of graphene devices, such as graphene p-n junctions.

We would like to acknowledge B. Wolfs, S. Bakker for technical
asistance. This work was financed by MSC$^{plus}$, NanoNed, NWO
(via a 'PIONIER' grant) and FOM.

\end{document}